\documentclass[twocolumn,trackchanges]{aastex63}
\hypersetup{linkcolor=red,citecolor=blue,filecolor=cyan,urlcolor=magenta}

\usepackage{enumerate}
\usepackage{amsmath}

\received{April XX, 2022}
\revised{??? xx, 2022}
\accepted{??? xx, 2022}
\submitjournal{AAS Journal}

\shorttitle{Observing SN Neutrinos. III. Parameter extraction}
\shortauthors{Suwa et al.}
\graphicspath{{./}{fig/}}

\begin{document}

\title{Observing Supernova Neutrino Light Curves with Super-Kamiokande.\\ 
III. Extraction of Mass and Radius of Neutron Stars from Synthetic Data}

\correspondingauthor{Yudai Suwa}
\email{suwa@yukawa.kyoto-u.ac.jp}

\author[0000-0002-7443-2215]{Yudai Suwa}
\affiliation{Department of Earth Science and Astronomy, The University of Tokyo, Tokyo 153-8902, Japan}
\affiliation{Center for Gravitational Physics and Quantum Information, Yukawa Institute for Theoretical Physics, Kyoto University, Kyoto 606-8502, Japan}

\author{Akira Harada}
\affiliation{Interdisciplinary Theoretical and Mathematical Sciences Program (iTHEMS), RIKEN, Wako, Saitama 351-0198, Japan}

\author{Masayuki Harada}
\affiliation{Department of Physics, Okayama University, Okayama 700-8530, Japan}

\author[0000-0003-0437-8505]{Yusuke Koshio}
\affiliation{Department of Physics, Okayama University, Okayama 700-8530, Japan}
\affiliation{Kavli Institute for the Physics and Mathematics of the Universe (Kavli IPMU, WPI), Todai Institutes for Advanced Study, \\ The University of Tokyo, Kashiwa 277-8583, Japan}

\author{Masamitsu Mori}
\affiliation{Department of Earth Science and Astronomy, The University of Tokyo, Tokyo 153-8902, Japan}

\author{Fumi Nakanishi}
\affiliation{Department of Physics, Okayama University, Okayama 700-8530, Japan}

\author[0000-0001-6330-1685]{Ken'ichiro Nakazato}
\affiliation{Faculty of Arts and Science, Kyushu University, Fukuoka 819-0395, Japan}

\author[0000-0002-9224-9449]{Kohsuke Sumiyoshi}
\affiliation{National Institute of Technology, Numazu College of Technology, Numazu 410-8501, Japan}

\author{Roger A. Wendell}
\affiliation{Department of Physics, Kyoto University, Kyoto 606-8502, Japan}
\affiliation{Kavli Institute for the Physics and Mathematics of the Universe (Kavli IPMU, WPI), Todai Institutes for Advanced Study, \\ The University of Tokyo, Kashiwa 277-8583, Japan}




\begin{abstract}

Neutrinos are guaranteed to be observable from the next galactic supernova (SN). Optical light and gravitational waves are also observable, but may be difficult to observe if the location of the SN in the galaxy or the details of the explosion are unsuitable. The key to observing the next supernova is to first use neutrinos to understand various physical quantities and then link them to other signals. In this paper, we present Monte Carlo sampling calculations of neutrino events from galactic supernova explosions observed with Super-Kamiokande. The analytical solution of neutrino emission, which represents the long-term evolution of neutrino-light curve from supernovae, is used as a theoretical template. It gives the event rate and event spectrum through inverse beta decay interactions with explicit model parameter dependence. Parameter estimation is performed on these simulated sample data by fitting least squares using the analytical solution. The results show that the mass, radius and total energy of a remnant neutron star produced by a SN can be determined with an accuracy of $\sim 0.1M_\odot$, $\sim 1$ km, and $\sim 10^{51}$ erg, respectively, for a galactic SN at 8 kpc.

\end{abstract}



\section{Introduction} 
\label{sec:intro}

The next Galactic supernova (SN) would provide a plethora of neutrinos which will likely be detected by a number of neutrino facilities such as Super-Kamiokande \citep{2012ARNPS..62...81S,2018JPhG...45d3002H}. 
While optical radiation and gravitational waves are also expected from supernova bursts, they may be difficult to observe if circumstances are unfavorable. 
For instance, any optical signals would be absorbed by the interstellar medium if the SN occurs near the Galactic anticenter. In addition, the expected amplitude of any gravitational wave from a SN is both highly uncertain and strongly model dependent, so it is unclear whether they can be detected for all types of explosions. 
Neutrinos on the other hand are strongly penetrating and known to be produced in abundance, making them a guaranteed signal that can be used to study the explosion and its mechanism.  
It will therefor be critical to connect this neutrino signal with other messengers from the next SN to extract the most information from the event.
In particular, understanding neutrinos at late times ($>$ 1 sec after the onset of explosion) is essential, since the physics surrounding this timescale has much smaller theoretical uncertainties than that of earlier times. 

Recently, late-time supernova neutrino emission has received much attention. \cite{2013ApJS..205....2N}, for example, 
conducted systematic protoneutron star (PNS) cooling simulations starting from the iron core's  collapse up to 20~s after the explosion. 
Further, \cite{2019ApJ...881..139S}, conducted simulations with the same method, but simulated beyond 100~s and investigated a data analysis method called {\it backward time analysis}. 
The work of \cite{2012PhRvL.108f1103R} presented similar simulations, while also accounting for convection inside the PNS \citep[see also][]{2021PhRvD.103b3016L}.
In addition to the PNS cooling simulations described above, there are also hydrodynamics simulations~\citep{2010A&A...517A..80F,2010PhRvL.104y1101H,2016NCimR..39....1M,2021PTEP.2021b3E01M}. 
However, these simulations are spherically symmetric (ie. one-dimensional, 1D) for the long-term evolution ($\gtrsim 10$ s) and require special initial conditions or artificial treatment to produce an explosion\citep[but see][in which a 2D simulation was used for a self-consistent explosion followed by a 1D simulation of the evolution up to 67 s]{2014PASJ...66L...1S}. 
More recent computational and technical advances allow for relatively long ($\gtrsim 1$ s) yet still systematic simulations in 2D or 3D \citep{2019PASJ...71...98N,2019MNRAS.489.2227V,2021Natur.589...29B,2021ApJ...915...28B,2021MNRAS.506.1462N}.
With these simulations multidimensional effects such as convection, standing-accretion shock instabilities, and lepton-number emission self-sustained asymmetries are naturally included in the neutrino signal. 
However, computational constraints nonetheless continue to limit the extension of these studies to later times.

Neutrino observations are sensitive to the physical quantities that characterize the central compact object left behind by a SN explosion.
For instance, the gravitational binding energy of the SN1987A event was estimated using its neutrino signal \citep{1987PhLB..196..267S,1988ApJ...334..891B}.
Since that binding energy, $E_{\rm b}$, is related to both the mass, $M$, and radius, $R$, of the PNS via $E_{\rm b}\sim M^2 R^{-1}$, it is difficult to make independent estimates of all three parameters. 
However, independent estimates of $M$ and $R$ would provide a deep understanding of nuclear physics through the nuclear equation of state \citep[see e.g.][]{2014EPJA...50...46F}. 
Therefore, we should aim to use neutrinos from the next Galactic supernova to determine $M$ and $R$ independently. 
Since it is unknown whether neutrinos alone can achieve this goal, in this paper we analyze simulated neutrino spectra to understand the limits of estimating these parameters.   

This paper is arranged as follows. Section \ref{sec:mock} describes how Monte Carlo calculations are used to create mock samples and Section \ref{sec:chi2} describes how the generated mock samples are analyzed. Section \ref{sec:error} describes the error evaluation from the data analysis method, and Section \ref{sec:summary} summarizes the main results.

\section{Mock sampling}
\label{sec:mock}

In this work we use the solution for the neutrino-light curve derived in \citet{2021PTEP.2021a3E01S}, in which the time evolution of the event rate and positron average energy are given by analytic functions of time. 
The parameter dependence on the mass and radius of the neutron star are also explicitly presented.

The event rate is given by
\begin{align}
\mathcal{R}&
=720\,{\rm s^{-1}}
\left(\frac{M_{\rm det}}{32.5\,{\rm kton}}\right)
\left(\frac{D}{10\,{\rm kpc}}\right)^{-2}
\nonumber\\
&\times
\left(\frac{M_{\rm PNS}}{1.4\,M_\odot}\right)^{15/2}
\left(\frac{R_{\rm PNS}}{10\,{\rm km}}\right)^{-8}
\left(\frac{g\beta}{3}\right)^{5}\nonumber\\
&\times\left(\frac{t+t_0}{100\,{\rm s}}\right)^{-15/2},
\label{eq:detection_rate}
\end{align}
where 
$M_{\rm det}$ is the detector mass and 32.5~kton corresponds to the entire inner detector volume of Super-Kamiokande,  $D$ is the distance between the SN and the Earth, $M_{\rm PNS}$ is the PNS mass,  $R_{\rm PNS}$ is the PNS radius,\footnote{This corresponds to the radius after PNS contraction by neutrino cooling.} $g$ is the density correction factor, and $\beta$ is the opacity boosting factor from coherent scattering \citep[see][for details]{2021PTEP.2021a3E01S}. 
The timescale $t_0$ is given by
\begin{align}
t_0&=210\,{\rm s}
\left(\frac{M_{\rm PNS}}{1.4\,M_\odot}\right)^{6/5}
\left(\frac{R_{\rm PNS}}{10\,{\rm km}}\right)^{-6/5}\nonumber\\
&\times\left(\frac{g\beta}{3}\right)^{4/5}
\left(\frac{E_{\rm tot}}{10^{52}\,{\rm erg}}\right)^{-1/5},
\label{eq:t0}
\end{align}
where $E_{\rm tot}$ is the total energy emitted by all flavors of neutrinos.
By integrating Eq. \eqref{eq:detection_rate} with Eq. \eqref{eq:t0}, 
the expected total number of events is
\begin{align}
    N&=\int_0^{\infty} \mathcal{R}(t)dt\nonumber\\
    &=89\left(\frac{M_{\rm det}}{32.5\,{\rm kton}}\right)
\left(\frac{D}{10\,{\rm kpc}}\right)^{-2}
\left(\frac{M_{\rm PNS}}{1.4\,M_\odot}\right)^{-3/10}\nonumber\\
&\times
\left(\frac{R_{\rm PNS}}{10\,{\rm km}}\right)^{-1/5}
\left(\frac{g\beta}{3}\right)^{-1/5}
\left(\frac{E_{\rm tot}}{10^{52}\,{\rm erg}}\right)^{13/10}.
\label{eq:total_number}
\end{align}
For the canonical parameters used in this paper ($M_{\rm det}=32.5$ kton, $D=8$ kpc, $M_{\rm PNS}=1.52 M_\odot$, $R_{\rm PNS}=11.8$ km, $g\beta=1.6$, and $E_{\rm tot}=10^{53}$ erg), the expectation becomes $N=2970$.

The average energy of created positrons is given by
\begin{align}
E_{e^+}
&=25.3\, {\rm MeV}
\left(\frac{M_{\rm PNS}}{1.4\,M_\odot}\right)^{3/2}\nonumber\\
&\times
\left(\frac{R_{\rm PNS}}{10\,{\rm km}}\right)^{-2}
\left(\frac{g\beta}{3}\right)
\left(\frac{t+t_0}{100\,{\rm s}}\right)^{-3/2}.
\label{eq:detection_energy}
\end{align}
For the energy distribution, we employ the Fermi-Dirac distribution function for the neutrinos, which allows us to calculate the distribution of positron as given in Appendix \ref{sec:appendix}. Note that the analysis shown in this paper uses only the average energy of Eq. \eqref{eq:detection_energy} and the spectrum information is not included.
As the results of \cite{2022ApJ...925...98N} indicate measurements of the average positron energy are largely insensitive to the details of the neutrino energy distribution, this choice is not expected to impact the results of the analysis below. 

\begin{figure}[tbp]
\centering
\includegraphics[width=0.45\textwidth]{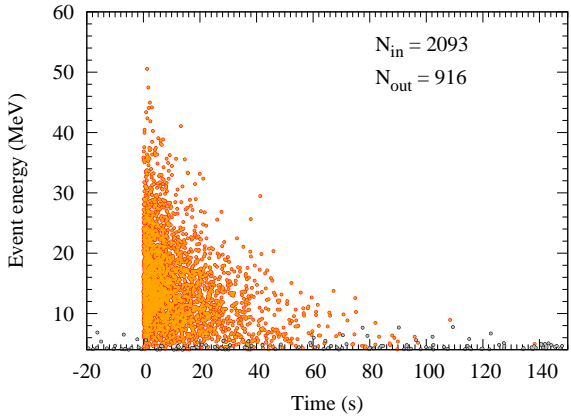}
\caption{An example scatter plot positron energy as function of time for a single simulation. The orange points indicate the signal and the gray points indicate the background. In this particular case, there are 3009 events detected, where 2093 are within the fiducial volume (22.5 kton) and 916 are not therein.}
\label{fig:mock}
\end{figure}

Using these equations, we simulate 100 Monte Carlo realizations of a SN with our canonical parameters.
Fig. \ref{fig:mock} shows an example.
We also plot the expected detector background modeled using \cite{2021mori}. This background is not otherwise used in the following analysis but will be discussed in the forthcoming paper. 
Note that the neutrino detection lasts for about 100~s, far beyond the timescale of current simulations using detailed treatments of neutrino-radiation hydrodynamics.
Accordingly, the analytic neutrino light curve is useful for filling this timescale gap.

\section{Data analysis}
\label{sec:chi2}

In this paper we use a least squares fit to estimate the SN parameters from the simulated data. 
The $\chi^2$ is given by
\begin{align}
    \chi^2=\sum_{i=1}^N \frac{(O_i-X_i)^2}{\sigma_i^2},
    \label{eq:chi2}
\end{align}
where $O_i$, $X_i$, and $\sigma_i$ are the observed value, the expected value, and the variance, respectively, with time index $i$. 
For the event rate ($X={\cal R}$), we use $\sigma_i^2={\cal R}_i^2/N_i$, where $N_i$ is the event number in the $i$-th bin.
For the average energy ($X=E_{e^+}$), we use $\sigma_i^2=(0.05E_{e^+})^2$ \citep{2022ApJ...925...98N}. 
We calculate the joint probability density function (PDF) for the measured parameters as
\begin{align}
    \mathcal{P}(M_{\rm PNS},R_{\rm PNS}, E_{\rm tot})\propto e^{-\chi^2(M_{\rm PNS},R_{\rm PNS}, E_{\rm tot})/2}.
    \label{eq:PDF}
\end{align}

\begin{figure}[tbp]
\centering
\includegraphics[width=0.45\textwidth]{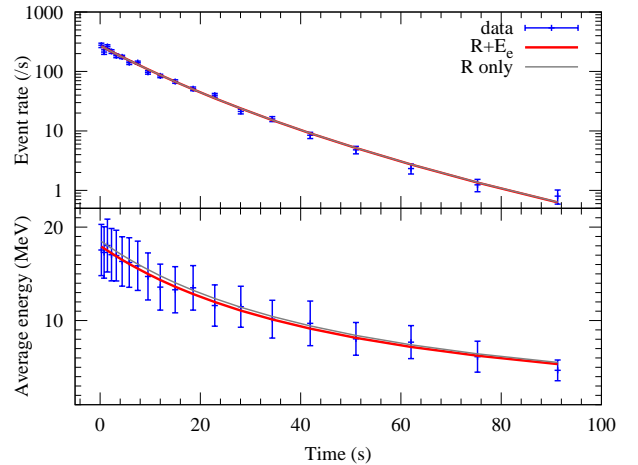}
\caption{Time evolution of the event rate (top) and average energy (bottom) from the simulation in Fig.~\ref{fig:mock}. The error bar on the event rate is the Poisson statistical error, while that for the average energy the error is 
taken from the standard deviation of the energies observed in that bin. Data is shown by blue points with error bars, the best fit models using the event rate alone is shown by the grey line, and the red line shows the fit using both the rate and average energy.}
\label{fig:rate_energy}
\end{figure}

\begin{figure}[tbp]
\centering
\includegraphics[width=0.45\textwidth]{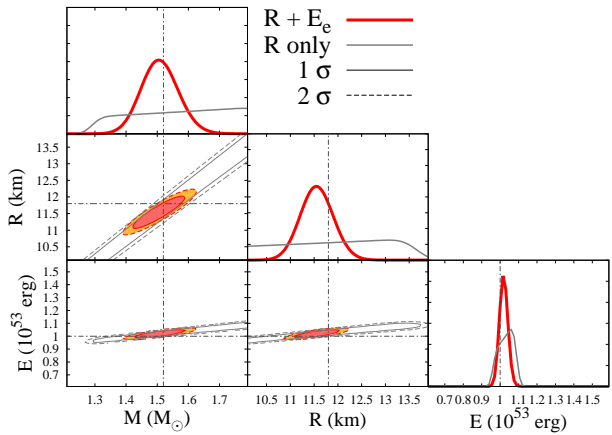}
\caption{Probability density function (PDF) determined by Eq. \eqref{eq:PDF} for the simulation in Fig.~\ref{fig:rate_energy}.
Contours with solid and dashed lines correspond to $\mathcal{P}/\mathcal{P}_{\rm max}=1/e$ (0.368, corresponding to $1\sigma$) and $1/e^2$ (0.135, $2\sigma$), respectively, where $\mathcal{P}_{\rm max}$ is the maximum value of the PDF. 
Colored lines give results using both the event rate and average energy of the positrons, while grey lines give those from the fit with the event rate alone.}
\label{fig:chi2}
\end{figure}

Fig. \ref{fig:rate_energy} shows the time evolution of the event rate and average energy of the model shown in Fig. \ref{fig:mock}. 
The time bins are calculated by
\begin{align}
    t_i&=t_{i-1} + \Delta t_i,\\
    \Delta t_i&= A\Delta t_{i-1},
\end{align}
where $\Delta t_i$ is the time width of $i$-th time bin and $A$ is a constant, which is estimated using the first time bin $t_1$ and last time bin of the analysis $t_{\rm end}$.
In this paper, we use $\Delta t_1=0.5$ and $t_{\rm end}=100$ s, respectively, and the number of bins $N=20$, such that $A\approx 1.206$. 
To calculate the $\chi^2$ in Eq. \eqref{eq:chi2} the central value of the time bin ($t_i-0.5\Delta t_i$) is used as shown in Figure \ref{fig:rate_energy}.
Error bars in the figure indicate the Poisson statistical error on the event rate and the standard deviation of energies in each bin for the average energy plot. 
Two solid lines (red and grey) show the time evolution of the event rate and average energy using the best fit parameters. The gray line is calculated by the event rate alone, but the red line is calculated by both the event rate and the average energy data.  For the latter the $\chi^{2}$ is estimated by the simple sum of the contributions at each time step from the event rate and average energy.

We next estimate the uncertainties of the model parameters with the PDF of Eq. \eqref{eq:PDF}. 
Figure \ref{fig:chi2} shows the distribution of $\mathcal{P}$. 
In this figure contours with solid and dashed lines correspond to $\mathcal{P}/\mathcal{P}_{\rm max}=1/e$ (0.368, corresponding to 1$\sigma$) and $1/e^2$ (0.135, $2\sigma$), respectively, where $\mathcal{P}_{\rm max}$ is the maximum value of $\mathcal{P}$. 
Colored lines give results with both $\mathcal{R}$ and $E_{e^+}$, but gray lines give those with $\mathcal{R}$ alone. 
Due to the parameter degeneracy, the allowed regions in $M$ and $R$ are rather large and elongated for the case with $\mathcal{R}$ alone. 
However, $E_{\rm tot}$ can be estimated precisely even using only  $\mathcal{R}$. 
We can draw the same conclusion from the marginalized PDFs shown in the top panel of each column.
Note that the uncertainties shown here are for a single realization. 
Since the observed data are subject to fluctuations due to Poisson statistics, it is necessary to perform Monte Carlo calculations for multiple realizations in order to evaluate the expected parameter sensitivity (expected error) in anticipation of an actual observation.

\section{Parameter Sensitivity}
\label{sec:error}

\begin{figure*}[htbp]
\centering
\includegraphics[width=0.8\textwidth]{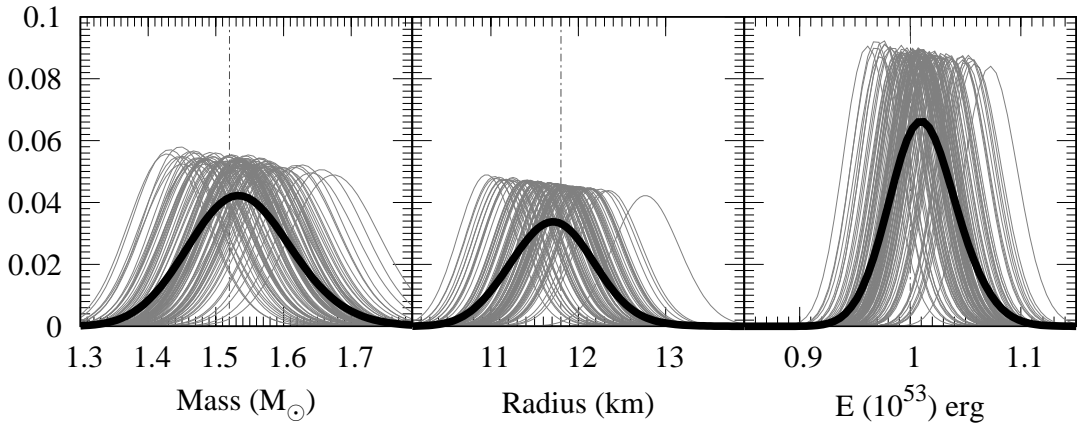}
\caption{PNS parameter PDFs, $M_{\rm PNS}$, $R_{\rm PNS}$, and $E_{\rm tot}$, respectively, for 100 realizations. }
\label{fig:pdf}
\end{figure*}

The expected parameter sensitivity is evaluated using 100 realizations of the model above.
Figure \ref{fig:pdf} shows the resulting PDFs of $M_{\rm PNS}$, $R_{\rm PNS}$, and $E_{\rm tot}$, respectively, for each realization.
The bold lines indicate their averages. 
Each Monte Carlo calculation produces various best-fit values according to Poisson statistics, but the average line shows that the input values are the most plausible and that values far from the input result in a predominant drop in the probability density.
Note that the average PDF taken over 1,000 realizations shows almost the same distribution.
Our estimate of the expected parameter sensitivity 
uses the median value of the average PDF as the central value and 68\% and 95\% uncertainties given by the parameter range representing those probability contents (taken about the median).
The results are summarized in Table \ref{tab:table}.

\begin{table}[tbp]
    \centering
    \caption{Expected values and statistical errors}
    \begin{tabular}{cccc}
        &  Median & 68\% & 95\% \\
    \hline
     $M_{\rm PNS}$ ($M_\odot$) &  1.532 & $^{+0.079}_{-0.075}$ & $^{+0.163}_{-0.147}$\\
     $R_{\rm PNS}$ (km) &  11.69 & $^{+0.48}_{-0.48}$ & $^{+0.98}_{-0.93}$\\
     $E_{\rm tot}$ ($10^{53}$ erg) &  $1.009$ & $^{+0.032}_{-0.030}$  &  $^{+0.066}_{-0.059}$\\ 
    \hline
    \end{tabular}
    \label{tab:table}
\end{table}

\begin{figure}[htbp]
\centering
\includegraphics[width=0.45\textwidth]{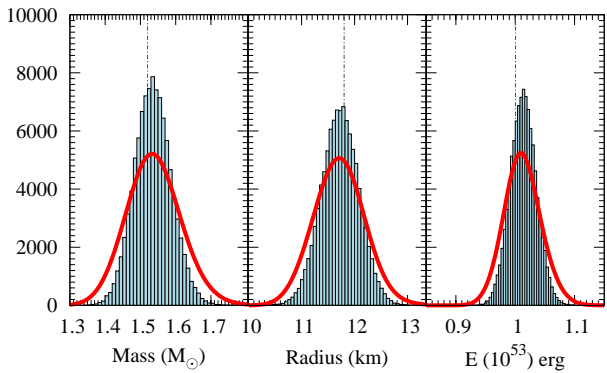}
\caption{Comparison of the distributions of the best-fit values of the PNS parameters (histogram) and their average PDFs (solid line). 
The histogram is drawn with 100,000 realizations
and the solid lines with 100.}
\label{fig:hist}
\end{figure}

Figure \ref{fig:hist} shows a histogram of the distribution of best-fit values for 100,000 realizations, with the average PDF shown as a red line. From this figure, we can see that the average PDF of 100 realizations is roughly equivalent to the best-fit fluctuations of the larger statistics. 
The slight deviation of the maximum of the best-fit value distribution from the input suggests that the fitting contains some bias. 
Since the goal of this paper is to demonstrate a simple model analyzed in a simple manner, we will not go into depth here.

Here we have only demonstrated the expected sensitivity 
for a specific choice of $M_{\rm PNS}$, $R_{\rm PNS}$, and $E_{\rm tot}$. 
Calculations performed for $1.2M_\odot<M_{\rm PNS}<1.8M_\odot$ and 10 km$<R_{\rm PNS}<$14 km,\footnote{$E_{\rm tot}$ is not changed since it is strongly related to the early phase.} however,
yield similar parameter uncertainties.
This can be understood from the limited dependence of $M_{\rm PNS}$ and $R_{\rm PNS}$ on the total number of events, as given by Equation \eqref{eq:total_number}.

\section{Summary}
\label{sec:summary}

In this paper, we performed Monte Carlo sampling calculations of neutrino events observed in Super-Kamiokande from a Galactic supernova explosion. 
The analytical solution from \cite{2021PTEP.2021a3E01S} was used as the theoretical template for the event rate and event spectrum from inverse beta decay interactions. 
Model parameter estimation was performed on these mock sampling data using a simple least-squares method.
The results show that the mass, radius and total energy can be determined with an accuracy of $\sim 0.1M_\odot$, $\sim 1$ km, and $\sim 10^{51}$ erg, respectively, for a Galactic SN at 8 kpc.

 We have focused specifically on the late neutrino emission from supernovae, since it depends on only a few physical quantities, such as the mass and radius of the neutron star. Importantly, both of these are reflected in the neutrino signal. 
 However, since the present method does not include the early-phase emission, the total event energy cannot be in this context~\cite[see. e.g.][for early emission]{2022MNRAS.512.2806N}. 
 The next step is to establish a methodology to approach the physical quantities of both the late and early neutrino emission. 
 This requires the development of a method to analyze numerical calculations of the complete neutrino emission over all time domains. The methodology presented in this paper is the first step toward this goal.

\acknowledgments

This work is supported by Grant-in-Aid for Scientific Research (20H00174, 20H01904, 20K03973) and Grant-in-Aid for Scientific Research on Innovative Areas (17H06357, 17H06365, 18H05437, 19H05811, 20H04747, 22H04571) from the Ministry of Education, Culture, Sports, Science and Technology (MEXT), Japan.
This work was partly supported by MEXT as ``Program for Promoting Researches on the Supercomputer Fugaku'' (Toward a unified view of the universe: from large scale structures to planets).

\appendix

\section{Energy distribution of positrons}
\label{sec:appendix}

We employ the Fermi-Dirac distribution for the neutrinos
and the resultant positron energy spectrum is given as
\begin{align}
    f(E)\propto \frac{E^4}{1+\exp(E/k_{\rm B}T_\nu)},
\end{align}
where $k_{\rm B}$ is the Boltzmann constant and $T_\nu$ is the temperature characterizing the intrinsic neutrino spectrum. 
Note that $T_\nu$ evolves with time as the PNS cools.
This spectrum is derived from the thermal neutrino spectrum ($\propto E^2/(1+\exp(E/k_{\rm B}T))$) and using  the inverse beta cross section ($\propto E^2$). 
It relates to Eq. \eqref{eq:detection_energy} as $E_{e^+}=\int Ef(E) dE/\int f(E) dE$ \cite[see][for details]{2021PTEP.2021a3E01S}.

\bibliography{nuLCmock}{}

\begin{thebibliography}{}
\expandafter\ifx\csname natexlab\endcsname\relax\def\natexlab#1{#1}\fi
\providecommand{\url}[1]{\href{#1}{#1}}
\providecommand{\dodoi}[1]{doi:~\href{http://doi.org/#1}{\nolinkurl{#1}}}
\providecommand{\doeprint}[1]{\href{http://ascl.net/#1}{\nolinkurl{http://ascl.net/#1}}}
\providecommand{\doarXiv}[1]{\href{https://arxiv.org/abs/#1}{\nolinkurl{https://arxiv.org/abs/#1}}}

\bibitem[{{Bollig} {et~al.}(2021){Bollig}, {Yadav}, {Kresse}, {Janka},
  {M{\"u}ller}, \& {Heger}}]{2021ApJ...915...28B}
{Bollig}, R., {Yadav}, N., {Kresse}, D., {et~al.} 2021, The Astrophysical
  Journal, 915, 28, \dodoi{10.3847/1538-4357/abf82e}

\bibitem[{{Burrows}(1988)}]{1988ApJ...334..891B}
{Burrows}, A. 1988, The Astrophysical Journal, 334, 891, \dodoi{10.1086/166885}

\bibitem[{{Burrows} \& {Vartanyan}(2021)}]{2021Natur.589...29B}
{Burrows}, A., \& {Vartanyan}, D. 2021, Nature, 589, 29,
  \dodoi{10.1038/s41586-020-03059-w}

\bibitem[{{Fischer} {et~al.}(2014){Fischer}, {Hempel}, {Sagert}, {Suwa}, \&
  {Schaffner-Bielich}}]{2014EPJA...50...46F}
{Fischer}, T., {Hempel}, M., {Sagert}, I., {Suwa}, Y., \& {Schaffner-Bielich},
  J. 2014, European Physical Journal A, 50, 46,
  \dodoi{10.1140/epja/i2014-14046-5}

\bibitem[{{Fischer} {et~al.}(2010){Fischer}, {Whitehouse}, {Mezzacappa},
  {Thielemann}, \& {Liebend{\"o}rfer}}]{2010A&A...517A..80F}
{Fischer}, T., {Whitehouse}, S.~C., {Mezzacappa}, A., {Thielemann}, F.~K., \&
  {Liebend{\"o}rfer}, M. 2010, Astronomy and Astrophysics, 517, A80,
  \dodoi{10.1051/0004-6361/200913106}

\bibitem[{{Horiuchi} \& {Kneller}(2018)}]{2018JPhG...45d3002H}
{Horiuchi}, S., \& {Kneller}, J.~P. 2018, Journal of Physics G Nuclear Physics,
  45, 043002, \dodoi{10.1088/1361-6471/aaa90a}

\bibitem[{{H{\"u}depohl} {et~al.}(2010){H{\"u}depohl}, {M{\"u}ller}, {Janka},
  {Marek}, \& {Raffelt}}]{2010PhRvL.104y1101H}
{H{\"u}depohl}, L., {M{\"u}ller}, B., {Janka}, H.~T., {Marek}, A., \&
  {Raffelt}, G.~G. 2010, Physical Review Letters, 104, 251101,
  \dodoi{10.1103/PhysRevLett.104.251101}

\bibitem[{{Li} {et~al.}(2021){Li}, {Roberts}, \&
  {Beacom}}]{2021PhRvD.103b3016L}
{Li}, S.~W., {Roberts}, L.~F., \& {Beacom}, J.~F. 2021, Physical Review D, 103,
  023016, \dodoi{10.1103/PhysRevD.103.023016}

\bibitem[{{Mirizzi} {et~al.}(2016){Mirizzi}, {Tamborra}, {Janka}, {Saviano},
  {Scholberg}, {Bollig}, {H{\"u}depohl}, \&
  {Chakraborty}}]{2016NCimR..39....1M}
{Mirizzi}, A., {Tamborra}, I., {Janka}, H.~T., {et~al.} 2016, Nuovo Cimento
  Rivista Serie, 39, 1, \dodoi{10.1393/ncr/i2016-10120-8}

\bibitem[{Mori(2021)}]{2021mori}
Mori, M. 2021, PhD thesis, Kyoto University

\bibitem[{{Mori} {et~al.}(2021){Mori}, {Suwa}, {Nakazato}, {Sumiyoshi},
  {Harada}, {Harada}, {Koshio}, \& {Wendell}}]{2021PTEP.2021b3E01M}
{Mori}, M., {Suwa}, Y., {Nakazato}, K., {et~al.} 2021, Progress of Theoretical
  and Experimental Physics, 2021, 023E01, \dodoi{10.1093/ptep/ptaa185}

\bibitem[{{Nagakura} {et~al.}(2021){Nagakura}, {Burrows}, \&
  {Vartanyan}}]{2021MNRAS.506.1462N}
{Nagakura}, H., {Burrows}, A., \& {Vartanyan}, D. 2021, Monthly Notices of the
  Royal Astronomical Society, 506, 1462, \dodoi{10.1093/mnras/stab1785}

\bibitem[{{Nagakura} \& {Vartanyan}(2022)}]{2022MNRAS.512.2806N}
{Nagakura}, H., \& {Vartanyan}, D. 2022, Monthly Notices of the Royal
  Astronomical Society, 512, 2806, \dodoi{10.1093/mnras/stac383}

\bibitem[{{Nakamura} {et~al.}(2019){Nakamura}, {Takiwaki}, \&
  {Kotake}}]{2019PASJ...71...98N}
{Nakamura}, K., {Takiwaki}, T., \& {Kotake}, K. 2019, Publications of the
  Astronomical Society of Japan, 71, 98, \dodoi{10.1093/pasj/psz080}

\bibitem[{{Nakazato} {et~al.}(2013){Nakazato}, {Sumiyoshi}, {Suzuki}, {Totani},
  {Umeda}, \& {Yamada}}]{2013ApJS..205....2N}
{Nakazato}, K., {Sumiyoshi}, K., {Suzuki}, H., {et~al.} 2013, The Astrophysical
  Journal Supplement Series, 205, 2, \dodoi{10.1088/0067-0049/205/1/2}

\bibitem[{{Nakazato} {et~al.}(2022){Nakazato}, {Nakanishi}, {Harada}, {Koshio},
  {Suwa}, {Sumiyoshi}, {Harada}, {Mori}, \& {Wendell}}]{2022ApJ...925...98N}
{Nakazato}, K., {Nakanishi}, F., {Harada}, M., {et~al.} 2022, The Astrophysical
  Journal, 925, 98, \dodoi{10.3847/1538-4357/ac3ae2}

\bibitem[{{Roberts} {et~al.}(2012){Roberts}, {Shen}, {Cirigliano}, {Pons},
  {Reddy}, \& {Woosley}}]{2012PhRvL.108f1103R}
{Roberts}, L.~F., {Shen}, G., {Cirigliano}, V., {et~al.} 2012, Physical Review
  Letters, 108, 061103, \dodoi{10.1103/PhysRevLett.108.061103}

\bibitem[{{Sato} \& {Suzuki}(1987)}]{1987PhLB..196..267S}
{Sato}, K., \& {Suzuki}, H. 1987, Physics Letters B, 196, 267,
  \dodoi{10.1016/0370-2693(87)90728-3}

\bibitem[{{Scholberg}(2012)}]{2012ARNPS..62...81S}
{Scholberg}, K. 2012, Annual Review of Nuclear and Particle Science, 62, 81,
  \dodoi{10.1146/annurev-nucl-102711-095006}

\bibitem[{{Suwa}(2014)}]{2014PASJ...66L...1S}
{Suwa}, Y. 2014, Publications of the Astronomical Society of Japan, 66, L1,
  \dodoi{10.1093/pasj/pst030}

\bibitem[{{Suwa} {et~al.}(2021){Suwa}, {Harada}, {Nakazato}, \&
  {Sumiyoshi}}]{2021PTEP.2021a3E01S}
{Suwa}, Y., {Harada}, A., {Nakazato}, K., \& {Sumiyoshi}, K. 2021, Progress of
  Theoretical and Experimental Physics, 2021, 013E01,
  \dodoi{10.1093/ptep/ptaa154}

\bibitem[{{Suwa} {et~al.}(2019){Suwa}, {Sumiyoshi}, {Nakazato}, {Takahira},
  {Koshio}, {Mori}, \& {Wendell}}]{2019ApJ...881..139S}
{Suwa}, Y., {Sumiyoshi}, K., {Nakazato}, K., {et~al.} 2019, The Astrophysical
  Journal, 881, 139, \dodoi{10.3847/1538-4357/ab2e05}

\bibitem[{{Vartanyan} {et~al.}(2019){Vartanyan}, {Burrows}, \&
  {Radice}}]{2019MNRAS.489.2227V}
{Vartanyan}, D., {Burrows}, A., \& {Radice}, D. 2019, Monthly Notices of the
  Royal Astronomical Society, 489, 2227, \dodoi{10.1093/mnras/stz2307}

\end{thebibliography}
\bibliographystyle{aasjournal}



\end{document}